\documentclass[12pt]{iopart}

\usepackage{graphicx}
\usepackage[square,sort&compress]{natbib}

\begin{document}

\article[Field-free molecular alignment probed by FLASH]{}{Field-free molecular alignment probed by the free electron laser in Hamburg (FLASH)}

\author{P Johnsson$^1$, A Rouz\'ee$^1$, W Siu$^1$, Y Huismans$^1$, F L\'epine$^2$, T Marchenko$^3$, S D\"usterer$^4$, F
Tavella$^4$, N Stojanovic$^4$, A Azima$^4$, R Treusch$^4$, M F Kling$^5$, M J J Vrakking$^1$}

\address{$^1$ FOM Institute for Atomic and Molecular Physics (AMOLF), Kruislaan 407, 1098 SJ Amsterdam, The Netherlands}
\address{$^2$ Universit\'e Lyon 1, CNRS, LASIM, UMR 5579, 43 bvd. du 11 novembre 1918, F-69622 Villeurbanne, France}
\address{$^3$ Laboratoire d'Optique Appliqu\'e, ENSTA/Ecole Polytechnique, Chemin de la Huni\`ere, 91761 Palaiseau, France}
\address{$^4$ Hamburger Synchrotronstrahlungslabor (HASYLAB) at Deutsches Elektronen-Synchrotron (DESY) Notkestrasse 85,
D-22603 Hamburg, Germany}
\address{$^5$ Max-Planck Institut f\"ur Quantenoptik, Hans-Kopfermann Strasse 1, D-85748 Garching, Germany}
\eads{\mailto{per.johnsson@fysik.lth.se}}

\begin{abstract}
We report experiments on field-free molecular alignment performed at FLASH, the free electron laser (FEL) in Hamburg. The impulsive alignment induced by a 100~fs near-infrared laser pulse in a rotationally cold CO$_{2}$ sample is characterized by ionizing and dissociating the molecules with a time delayed extreme ultra-violet (XUV) FEL pulse. The time-dependent angular distributions of ionic fragments measured by a velocity map imaging spectrometer shows rapid changes associated with the induced rotational dynamics. The experimental results also show hints of a dissociation process that depends non-linearly on the XUV intensity. With samples of aligned molecules at FLASH, experiments using ultrashort XUV pulses become possible in the molecular frame, which will enable new insights into the understanding of molecules and their interactions.
\end{abstract}

\pacs{41.60.Cr, 33.80.Gj, 33.80.Rv}
\maketitle

\section{\label{sec:intro}Introduction}
Coherent light from lasers is a powerful tool to explore the quantum world of atoms and molecules. Laser technology has matured rapidly during the second part of the last century and high power ultrashort radiation sources are now available. The most common ultrashort laser systems are based on Ti:Sapphire technology and provide radiation in the near-infrared (IR) regime with a pulse duration down to the few-cycle regime. These lasers allow to perform time-resolved pump-probe studies, where a first \textquotedblleft pump" pulse initiates dynamics in a system that is then investigated after a given delay by means of a second \textquotedblleft probe" pulse. Femtosecond pump-probe experiments have extensively been used to follow atomic and molecular motion in time~\cite{BaumertPRL1991,ZewailJPCA2000}. Furthermore, near-IR femtosecond technology has given birth to the field of attosecond science~\cite{AgostiniRPP2004, KlingARPC2008}, where attosecond extreme ultraviolet (XUV) pulses are obtained from high-order harmonics generated through strong-field interactions in atomic or molecular gases~\cite{FerrayJPB1988, DrescherScience2001, PaulScience2001}.

Although providing the sub-femtosecond resolution required to investigate ultrafast electron dynamics, the photon flux of the attosecond sources available today is still low ($10^{6}-10^{10}$~photons/pulse), making experiments requiring high intensities (such as those involving non-linear processes) or large signal strengths (like for single-shot imaging experiments) very challenging. Free electron lasers (FELs), such as the free electron laser in Hamburg (FLASH), have become interesting complementary sources, producing fluxes on the order of $10^{13}$~photons/pulse in the XUV region, with pulse durations of 10-15~fs~\cite{AyvazyanEPJD2006,AckermannNPhot2007}. A first glimpse of the possibilities offered by XUV FEL sources was obtained already in 2002 with the pre-cursor of FLASH, the TESLA Test Facility FEL. Wabnitz et al. illuminated xenon clusters with 12.7~eV radiation, focused to intensities of up to $10^{14}$~W/cm$^{2}$. They observed multiply charged ions up to Xe$^{8+}$ resulting from Coulomb explosion of the clusters after sequential absorption of on average 30 photons~\cite{WabnitzNature2002}. Ionization of single xenon atoms, producing multiply charged ions up to Xe$^{6+}$ by multiphoton sequential ionization was also carried out by Wabnitz et al.~\cite{WabnitzPRL2005}. In a more recent experiment, where FLASH (operating at a photon energy of 98 eV) was focused to intensities of $10^{16}$~W/cm$^{2}$, even higher degrees of ionization (up to Xe$^{21+}$) were observed~\cite{SorokinPRL2007}.

One of the main driving forces behind the development of high flux XUV sources like FLASH, and in particular future x-ray FELs, is the possibility of diffractive imaging of small objects~\cite{SchotteScience2003}. In a diffraction experiment structural information is encoded in interference patterns that result from the way that an electron or light wave scatters. In the case of light diffraction, the required wavelength is in the x-ray regime, and it has been proposed to use the future x-ray FELs to record single-shot diffraction ptterns from isolated molecules in order to retrieve their structure~\cite{NeutzeNature2000}. Already, during the first two years of operation of FLASH, very promising examples of this concept have been obtained at longer wavelengths~\cite{ChapmanNatPhys2006, BoganNL2008}.

Alternatively, an imaging experiment can rely on electron diffraction, using either external electron pulses~\cite{IheeScience2001,SiwickScience2003} or electrons created within the molecule, through x-ray photo-ionization. In the latter approach, where one \textquotedblleft illuminates the molecule from within" \cite{LandersPRL2001}, the electrons that are ejected from the molecule carry information about the molecular structure in a variety of ways. Electrons that are ejected from localized sites within a molecule (by atomic inner-shell ionization) encounter the other atoms in the molecule as scattering centres. The scattering processes are imprinted in the angular distributions of the ejected photoelectrons (provided these are measured in the molecular frame). If the photo-emission can occur from multiple equivalent atoms within the molecule, additional interference structures can be observed, since one cannot distinguish from which atom an electron is ejected.

So far, a few pioneering experiments have been performed that illustrate the possibility to obtain structural information by making use of energetic electrons that are created in the molecule itself. These studies have been performed at synchrotrons on ground state molecules that do not undergo time-dependent dynamics. The first example where the concept of \textquotedblleft illuminating a molecule from within" was applied was given by D\"orner and co-workers, who studied inner-shell photoionization of the CO molecule by exciting the molecule above the carbon K-edge~\cite{LandersPRL2001}. By measuring the angular distributions of the ejected photoelectrons in coincidence with the recoil momenta of C$^{+}$ and O$^{+}$ ionic fragments, they could reconstruct the molecular-frame photoelectron angular distributions (MF-PADs), thereby revealing the electron wave propagation in the 3D potential energy landscape of the molecule. The authors of this paper immediately realized the possibility of using this method as a way to probe photochemical reactions triggered by short laser pulses, but perceived as a problem that in a coincidence experiment an MF-PAD measurement requires a fragmentation process that yields ionic fragments with an axial recoil. This problem can be overcome by applying molecular alignment and orientation techniques, preparing an aligned (or oriented) sample of molecules so that any measurement done in the laboratory frame is automatically also done in the molecular frame. In the femtosecond/attosecond laser community, recent experiments have already demonstrated the potential of using a sample of field-free aligned molecules with a high degree of alignment. Benchmark experiments include the determination of the electronic~\cite{KanaiNature2005} and nuclear~\cite{ItataniNature2004} structure of molecules by means of high-order harmonic generation, as well as by electron tunneling and diffraction~\cite{MeckelScience2008}.

Here we report on the first demonstration where an aligned molecular sample is probed by the femtosecond XUV pulses from FLASH. A field-free sample of aligned molecules was prepared using the FLASH in-house Ti:Sapphire IR laser~\cite{RadcliffeNIMA2007, RedlinInPrep2009}, and the degree of alignment was subsequently measured by a velocity map imaging spectrometer, recording the momentum distribution of O$^{+}$ fragments resulting from Coulomb explosion of CO$_{2}^{2+}$ ions created through photo-ionization by 46~eV photons. Further, we have made a first attempt to access the MF-PADs by measuring the photoelectron momentum distribution from ionization of the aligned molecules, allowing us to make preliminary observations of the dependence of the PADs on the molecular alignment distribution. Bringing aligned molecular samples to the focus of FLASH opens the door to future work applying XUV photo-ionization, angle-resolved photoelectron spectroscopy and time-resolved dissociation studies in the molecular frame.

\section{\label{sec:setup}Experimental setup \& technique}
The FLASH facility has been described in detail elsewhere~\cite{AyvazyanEPJD2006,AckermannNPhot2007}, and here we only describe the conditions of the present study. The experiment was performed at the primary focus of BL2, where the focal spot size is $\sim15-25~\mu$m full-width half-maximum (FWHM). The FEL was operated in multibunch mode, with 30 bunches in a bunchtrain, a bunch separation of 10~$\mu$s and a bunch train repetition rate of 5~Hz.
In the present experiment, advantage could not be taken of the multibunch mode, and the measurements had to select a single pulse from the bunch train, as the milli-Joule part of the IR laser installation had a repetition rate of 5~Hz. The photon energy used for the experiment was 46~eV ($\lambda=27$~nm) and the average pulse duration was estimated to 15~fs. The average pulse energy was 50~$\mu$J and the rms deviation was 15~$\mu$J. In order to be able to extract the intensity-dependence from the recorded data, all data were acquired and saved on a single-shot basis together with the global bunch ID from the FLASH data acquisition system. After the experiment, the data could then be correlated and sorted according to the pulse energy, which was also recorded on a single-shot basis by the in-house gas monitor detectors (GMDs)~\cite{TiedkeJAP2008}.

\subsection{\label{ssec:alignment}Laser-induced molecular alignment}
This last decade, considerable attention has been devoted to laser-controlled alignment (for a review, see~\cite{StapelfeldtRMP2003} and references therein). A strong non-resonant laser pulse can induce a torque via the interaction of the laser electric field with the molecular polarizability resulting in an alignment of the molecular axis along the laser polarization. For pulse durations much larger than the molecular rotational period the photo-excitation leads to an adiabatic alignment during the laser pulse~\cite{FriedrichPRL1995, SakaiJCP1999, LarsenJCP1999}, while excitation by a shorter laser pulse yields a periodic alignment that occurs after the laser field is gone, when the induced rotational wave packet re-phases~\cite{SeidemanPRL1999, RoscaPrunaPRL2001}. While the adiabatic approach in general allows for higher degrees of alignment, the non-adiabatic or impulsive technique leads to alignment of the molecules under field-free conditions, which is a pre-requisite in experiments where the presence of an external field might perturb the system under study. The present study was performed under non-adiabatic conditions.

\begin{figure}[h]\centering
\includegraphics[width=1\linewidth]{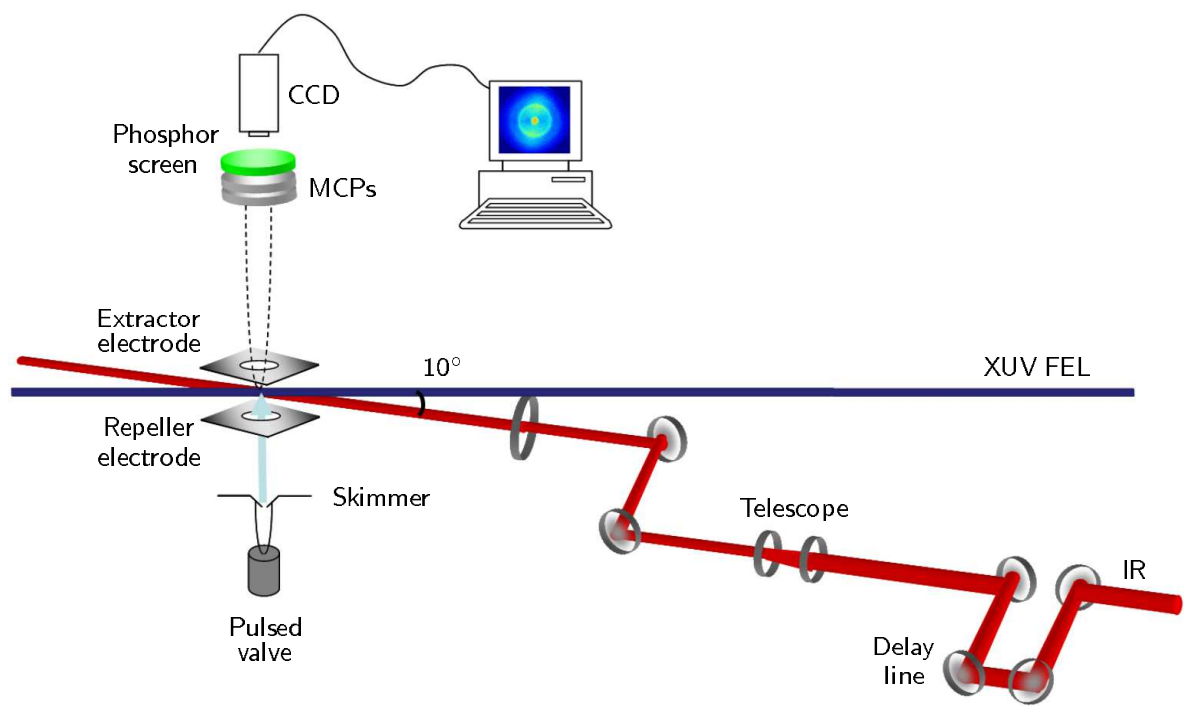}
\caption{\label{fig:setup}(Colour online) Schematic representation of the experimental setup. The two beams, propagating at an angle of $10^{\circ}$ to each other, intersect a molecular beam at the center of a velocity map imaging spectrometer. The near-IR beam is used to impulsively align the molecules that are probed by ionizing and dissociating them using the time-delayed FEL pulse. The degree of alignment is quantified using the 2D momentum distribution of the ionic fragments, which is observed by a detector imaged by a CCD camera.}
\end{figure}

\subsection{\label{ssec:pplaser}The FLASH pump-probe laser}
To align the molecules, the present experiment used the in-house IR laser at FLASH, described in detail in~\cite{RadcliffeNIMA2007, RedlinInPrep2009}. Briefly, this Ti:Sapphire laser system produced 100~fs laser pulses at a wavelength of 800~nm, with a pulse energy of a few mJ at a repetition rate of 5~Hz, synchronized with the pulse pattern of FLASH. The IR laser is located in a laser hutch, from which the pulses are transported under vacuum over a distance of almost 20~meters to the experimental station. The beam transport line includes a delay stage that can be used to vary the delay between the IR laser pulses and the XUV-FEL pulses. At the experiment, the IR beam size was reduced by means of a telescope and focused into the experimental chamber by a 300~mm focal length lens to obtain a spot size of 30-40~$\mu$m. For the present experiment, it was crucial to keep the focal spot size of the IR larger than that of the XUV, to ensure that all the molecules in the probe volume had been exposed to the alignment pulse. The overlap with the FEL was made in a non-collinear arrangement, with an angle of 10$^\circ$ between the two beams~(see \fref{fig:setup}).

\subsection{\label{ssec:overlap}Spatial and temporal overlap}
One of the most challenging aspects of doing a pump-probe experiment at FLASH is the spatial and temporal overlap between the IR and XUV-FEL pulses. Provided that the pulses are temporally overlapped, the spatial overlap can be obtained and optimized using a suitable two-color signal, like for example sideband generation in the photoelectron spectra~\cite{MeyerPRA2006}. However, as the temporal jitter between the FEL and the IR laser can be as large as 1~ps, this can be tedious work. In our case, the temporal and spatial overlap was obtained by observing the appearance of the bond-softening dissociation pathway in H$_{2}$ or D$_{2}$ molecules when the IR field arrives after the XUV pulse~\cite{JohnssonInPrep2009}. With this method, it is only required that the XUV pulse arrives before the IR pulse in order to observe the signal at spatial overlap. Once spatial overlap is achieved, it is straight-forward to scan the delay to find the temporal overlap. As will be shown later, the delay resolution of 1~ps is not good enough to properly resolve the re-phasing of the rotational wave packets created by the alignment pulse. However, a better temporal resolution could be achieved by using the data from the in-house timing electro-optical (TEO) sampling system which measured the exact delay between the IR laser pulse and the FEL pulse on a shot-to-shot basis~\cite{AzimaAPLSub2009}. As all data were recorded single-shot, this allowed for off-line sorting of the data with respect to the measured delays, leading to a final delay resolution below 100~fs.

\subsection{\label{ssec:vmis}The velocity map imaging spectrometer}
At the center of the velocity map imaging spectrometer (VMIS)~\cite{EppinkRSI1997}, the XUV and IR beams intersected a molecular beam formed by supersonic expansion in a pulsed gas valve operating at 5~Hz synchronized with the FEL. As shown in \Fref{fig:setup}, the molecular beam was perpendicular to the laser propagation direction and propagated towards the detector. The VMIS allows to record the 2D momentum distribution of ions or electrons resulting from photo-ionization/dissociation by the XUV FEL at very high count rates, making it an ideal technique for high flux sources like FLASH~\cite{JohnssonJMO2008}. In the VMIS the charged particles are accelerated by a static electric field formed by a repeller and an extractor electrode, and are then projected onto a detector composed of dual micro-channel plates (MCPs) and a phosphor screen, followed by a CCD camera. A 200~ns electrical gate was applied to the back of the MCPs to collect only the particles from a single FEL pulse in the bunch train, as well as allowing for mass selection when detecting ions. The 3D momentum distribution was reconstructed from the measured 2D projection using a mathematical inversion procedure~\cite{VrakkingRSISub2008}.

\begin{figure}[h]\centering
\includegraphics[width=1\linewidth]{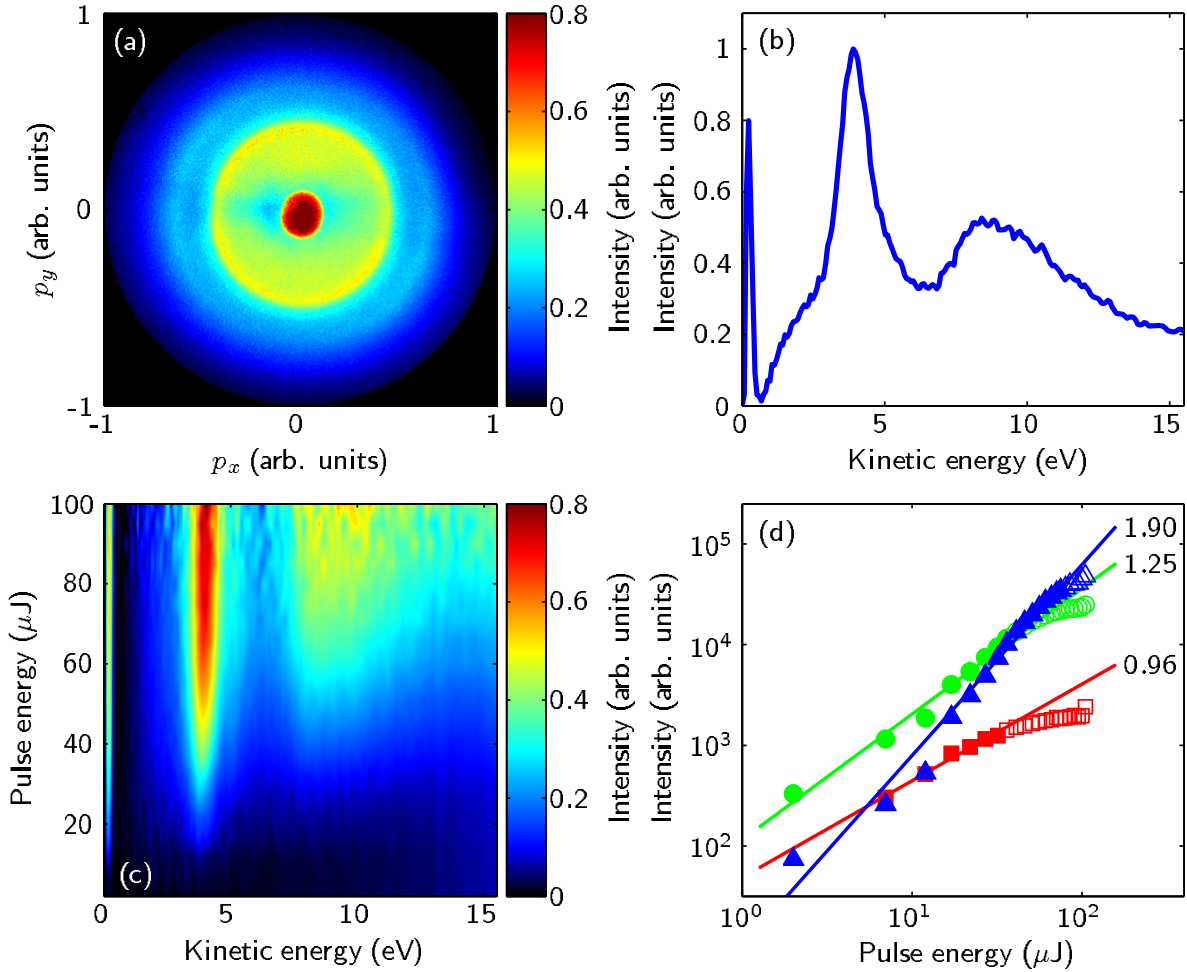}
\caption{\label{fig:Oplus}(Colour online) Experimentally measured O$^+$ 2D momentum distribution, averaged over 5000~shots, from dissociation of CO$_2$ by the FEL pulses alone (a) and the corresponding kinetic energy spectrum (b). The $p_y$-axis is parallel to the XUV polarization axis. Panel (c) shows the evolution of the O$^+$ kinetic energy spectrum with the FEL pulse energy while panel (d) shows the integrated signal in the different dissociation channels as a function of pulse energy (0-1~eV: red squares; 1-7~eV: green circles; 7-15~eV: blue triangles). The solid lines have been fitted in the non-saturated regime, indicated by filled symbols, and the slopes of the fitted lines are indicated in the plot.}
\end{figure}

\section{\label{sec:frag}XUV photo-dissociation of CO$_2$}
A typical 2D momentum distribution of O$^{+}$ fragments resulting from the dissociative ionization and Coulomb explosion of CO$_{2}$ molecules by the XUV FEL pulses alone is shown in \fref{fig:Oplus}(a). The $p_y$-axis is parallel to the XUV polarization axis. The data were recorded by averaging over 5000~FEL pulses. The corresponding kinetic energy spectrum is shown in \fref{fig:Oplus}(b). Three main features are readily observed, namely an intense peak below 0.5~eV, attributed to dissociative ionization (CO$_{2}^{+}\rightarrow$ CO + O$^{+}$), and two outer rings centered at 3.9~eV and 8.7~eV. The ring at 3.9~eV has been assigned in a previous study performed using the unfocused FEL beam (and thus at a much lower intensity) to the two-body fragmentation pathway CO$_{2}^{2+}\rightarrow$ CO$^{+}$ + O$^{+}$~\cite{JohnssonJMO2008}, while the high kinetic energy contribution was absent in that data.

\subsection{\label{ssec:xuvonly}Intensity dependence}
The dependence of the different fragmentation pathways on the FEL intensity can be obtained by sorting the single-shot data by the pulse intensity, as measured by the GMD (see \sref{sec:setup}). The result of this procedure is shown in \fref{fig:Oplus}(c). It is clear that the high kinetic energy contribution appears only for higher pulse energies. By comparing the non-linearity of the two channels [\fref{fig:Oplus}(d)], it can be concluded that the high energy channel involves the absorption of two XUV photons rather than a single photon (as for the lower energy channels). The total energy absorbed by the molecule is thus 92~eV. In a previous study, Masuoka \etal show that for photon energies up to 100~eV the maximum kinetic energy release in the two-body dissociation channel is below 10~eV, meaning a maximum kinetic energy of 6.4~eV for O$^+$~\cite{MasuokaJCP1996}. Thus, the observed high-energy fragments are likely to originate from one of the three-body dissociation channels listed below:

\begin{eqnarray}
\mathrm{CO}_2^{2+}&\rightarrow& \mathrm{C}^+ + \mathrm{O}^+ + \mathrm{O}   \\
\mathrm{CO}_2^{2+}&\rightarrow& \mathrm{C} + \mathrm{O}^+ + \mathrm{O}^+   \\
\mathrm{CO}_2^{3+}&\rightarrow& \mathrm{C}^+ + \mathrm{O}^+ + \mathrm{O}^+
\end{eqnarray}

\noindent
The fragmentation of doubly ionized CO$_2$ into C$^+$, O$^+$ and O (1), has an appearance energy of 47.2~eV, and is thus a possible dissociation pathway~\cite{MasuokaPRA1994}. For this channel, Masuoka \etal have observed maximum kinetic energies for O$^+$ up to 10~eV, and for C$^+$ fragments around 2~eV~\cite{MasuokaJCP1996}. The kinetic energies for the C$^+$ fragments are also in agreement with the observed kinetic energies in the current study (not shown here). The dissociation into two O$^+$ ions and a carbon neutral (2) has previously been observed at a photon energy of 60~eV~\cite{MillieJCP1985}. In a recent electron impact experiment, the O$^+$ fragment energies in this channel have been observed to be very similar to those of (1)~\cite{BapatJPB2007}. Finally, although triple ionization (3) is possible at energies above about 80~eV~\cite{MasuokaPRA1994}, it is a weak process, and we therefore expect the contribution of this fragmentation pathway to be negligible compared to the other two channels. In conclusion, we have strong indications of the presence of a dissociation process that depends non-linearly on the FEL intensity. Due to the density of states as well as the various possible fragmentation pathways, it is not possibly to quantify the contribution from the different channels. A fully correlated experiment using a COLTRIMS setup, would be able to answer this question.

\begin{figure}[h]\centering
\includegraphics[width=1\linewidth]{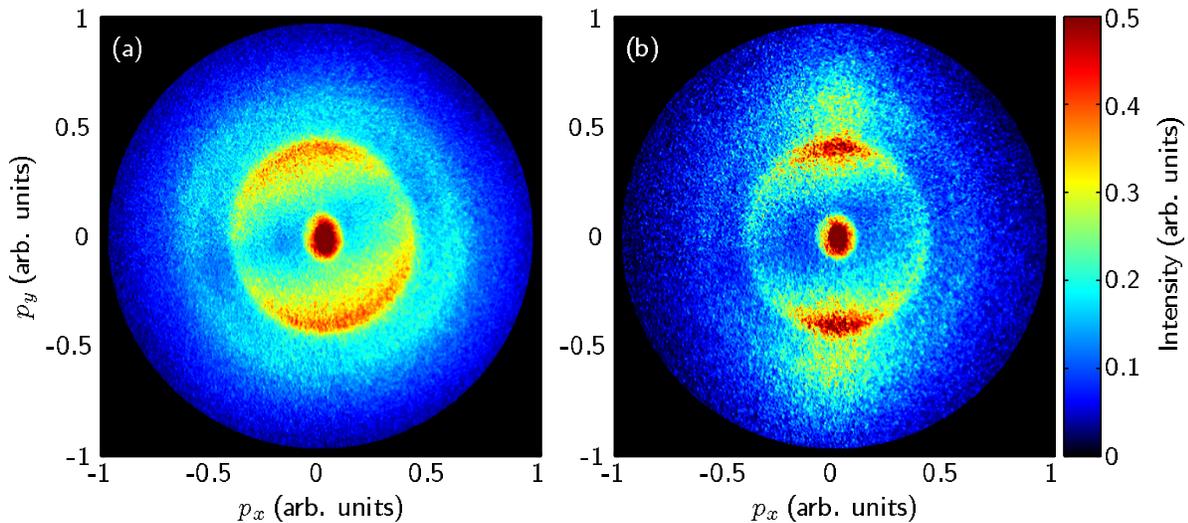}
\caption{\label{fig:Oplusaligned}(Colour online) Experimental O$^{+}$ 2D momentum distributions, averaged over 5000 shots, from dissociation of CO$_2$ by a combination of FEL and IR pulses. The two images are taken when the FEL pulse precedes the IR Field (a) and when the FEL and IR pulses overlap in time (b). For both panels, the $p_y$-axis is parallel to the IR polarization axis.}
\end{figure}

\subsection{\label{ssec:expalign}Molecular alignment}
In the following, the IR pulses were introduced and overlapped spatially with the XUV pulses. The IR intensity was kept sufficiently low, in order not to induce any significant ionization or dissociation by the alignment pulse alone. \Fref{fig:Oplusaligned} shows the O$^{+}$ momentum distributions recorded when the FEL pulse precedes the IR pulse (a) and at the temporal overlap of the IR and XUV pulses (b).
In this configuration, the $p_y$-axis was chosen to be parallel to the IR polarization axis. When these data were recorded, the IR pulse was stretched to a pulse duration of 1~ps in order to ensure some temporal overlap even with the 1~ps jitter. As seen, when the FEL pulse comes before the IR field, the angular distribution of the O$^{+}$ fragments is broad. When the pulses overlap in time, the O$^{+}$ fragments peak along the IR polarization axis, indicating that the molecules are now aligned preferentially along the polarization direction of the IR field at the time of ionization by the XUV, and that the dissociation in the Coulomb explosion channels proceeds rapidly enough so that the molecules do not have time to rotate further during the dissociation.

Molecular alignment is generally quantified by the averaged value $\langle\cos^{2}\theta\rangle$, where $\theta$ is the angle between the molecular axis and the IR polarization axis, and the averaging made over the 3D momentum distribution obtained after inversion of the recorded 2D data. When $\langle\cos^{2}\theta\rangle=1$, the molecules are fully aligned along the laser polarization axis whereas when $\langle\cos^{2}\theta\rangle=0$, molecules are in the plane perpendicular to the polarization axis (which is called \textquotedblleft anti-alignment" or planar de-localization). The value $\langle\cos^{2}\theta\rangle = 1/3$ corresponds to an isotropic molecular sample. For the present results, the $\langle\cos^{2}\theta\rangle$ value for the inner Coulomb explosion channel (1-7~eV), is 0.35 and 0.50 without and with the IR field at overlap, respectively. In principle, one has to be concerned with the possibility of both geometrical and dynamical alignment by the probe laser pulse. Geometric alignment refers to the dependence of the ionization and Coulomb explosion process on the angle between the molecular axis and the polarization axis of the FEL pulse (angular selectivity). As the angular distribution measured with the IR pulse absent is close to isotropic, the induced geometrical alignment in the present case is small, meaning that the observed angular distribution is largely determined by the molecular alignment of the CO$_{2}$ molecules by the IR pulse. This observation is consistent with an earlier study where it was shown that by using one-photon XUV ionization as the probe process (like for the inner Coulomb explosion channel in the present data), one decreases the influence of the probe laser electric field on the molecular alignment~\cite{LepineJMO2007}.

\begin{figure}[h]\centering
\includegraphics[width=1\linewidth]{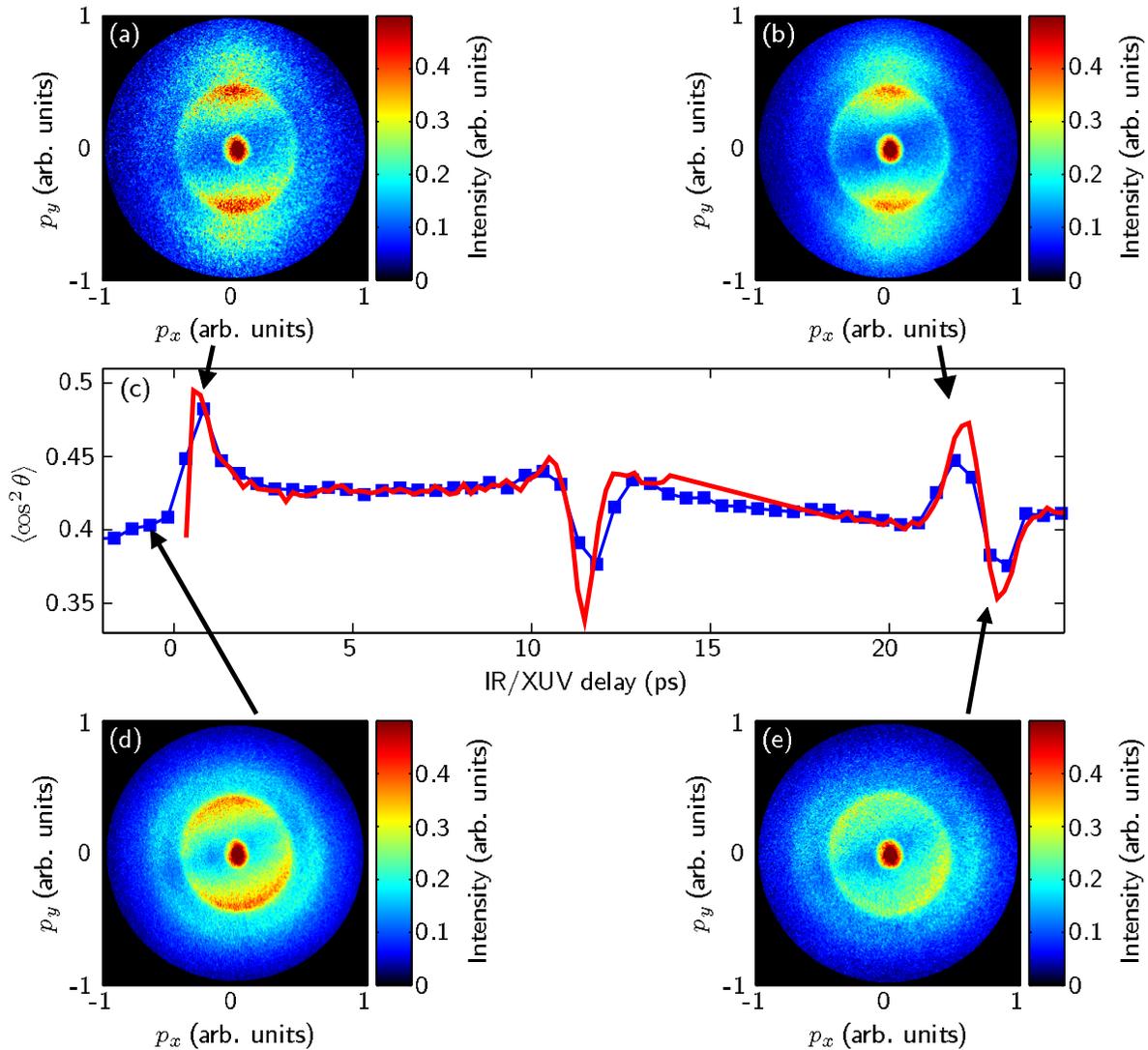}
\caption{\label{fig:revivals}(Colour online) Evolution of $\langle\cos^{2}\theta\rangle$ as a function of the delay between the alignment IR pulse and the XUV probe pulse (c). The blue squares show the values extracted from the experimentally averaged data (averaged over 1000~shots/delay), and the red line shows the values extracted from the TEO-sorted data. Panels (a), (b), (d) and (e) show the O$^+$ 2D momentum distributions obtained after sorting the data, for the different delays indicated by the arrows. The $p_y$-axis is parallel to the IR polarization axis. }
\end{figure}

\subsection{\label{ssec:expalignff}Field-free alignment}
Once the effect at temporal overlap was observed, the IR pulses were re-compressed to 100~fs. The O$^{+}$ fragment angular distribution was recorded with the pump-probe delay changing in steps of 0.5~ps in a range of -1 ps to 25 ps. The results are presented in \fref{fig:revivals}. The short IR excitation leads to the formation of a rotational wave packet by multiple two-photon Raman processes. The periodic re-phasing of this coherent rotational wave packet after the IR pulse has ended gives rise to revivals of the alignment at time delays that are determined by the rotational constant of the molecule. In the case of CO$_{2}$, the rotational constant is $B=0.3902$~cm$^{-1}$, giving a rotational period of $T_{r}=1/(2Bc)=42$~ps. Revivals of alignment are expected every $T=T_{r}/4$. \Fref{fig:revivals}(c) shows the evolution of $\langle\cos^{2}\theta\rangle$ as a function of the delay between the alignment IR pulse and the XUV probe pulse. When the XUV and the IR pulses overlap, the angular distribution starts to peak along the pump laser polarization direction, resulting in a higher value of $\langle\cos^{2}\theta\rangle$. The highest value of alignment is reached shortly after the time-overlap since the alignment dynamics cannot follow the short IR excitation. For longer pump-probe delays, rapid changes in $\langle\cos^{2}\theta\rangle$ are observed every 10.5~ps. This is a clear signature of the field-free alignment of the molecules along the IR field polarization. While the blue squares show the values extracted from the experimentally averaged data, the red line shows the values extracted after sorting the data according to the delays from TEO. It is clearly seen that by sorting the data the temporal resolution is improved, as the transients in $\langle\cos^{2}\theta\rangle$ are narrowed. The improved temporal resolution also leads to improved values of $\langle\cos^{2}\theta\rangle$, showing that without the sorting, the delay jitter averages out the degree of alignment. Panels (a), (b), (d) and (e) show O$^+$ 2D momentum distributions generated from the sorted data, for the different delays indicated by the arrows. The $p_y$-axis is parallel to the IR polarization axis. Images (d) and (a) were recorded before and after overlap, whereas panels (b) and (e) correspond to images recorded around the second revival, showing an alignment (b) and a planar delocalization (e). It should be noted that the maximum degree of alignment measured in this experiment ($\langle\cos^{2}\theta\rangle$=0.50) is relatively small. This can be partly attributed to the intensity of the IR field that was deliberately kept low to avoid any ionization.

\begin{figure}[h]\centering
\includegraphics[width=1\linewidth]{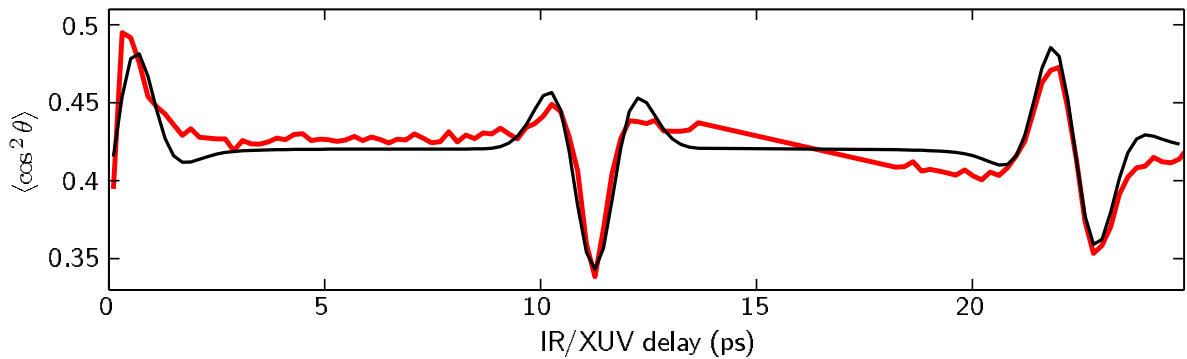}
\caption{\label{fig:theory}(Colour online) Comparison between experiment and theory. The red line shows the experimentally determined $\langle\cos^{2}\theta\rangle$, and the black line is a fit performed by solving the time-dependent Schr\"odinger equation.}
\end{figure}

\Fref{fig:theory} shows the result of a simulation where $\langle\cos^{2}\theta\rangle$ has been computed by solving the time-dependent Schr\"{o}dinger equation using the estimated intensity~\cite{SeidemanPRL1999}. A least squares fitting method was employed to find the set of parameters (IR intensity, rotational temperature of the molecules and the offset due to geometric alignment), that lead to the best agreement with the experimental curve. A very good agreement was found for a rotational temperature of 40~K and an intensity of $7\cdot10^{12}$~W/cm$^{2}$.

\begin{figure}[h]\centering
\includegraphics[width=1\linewidth]{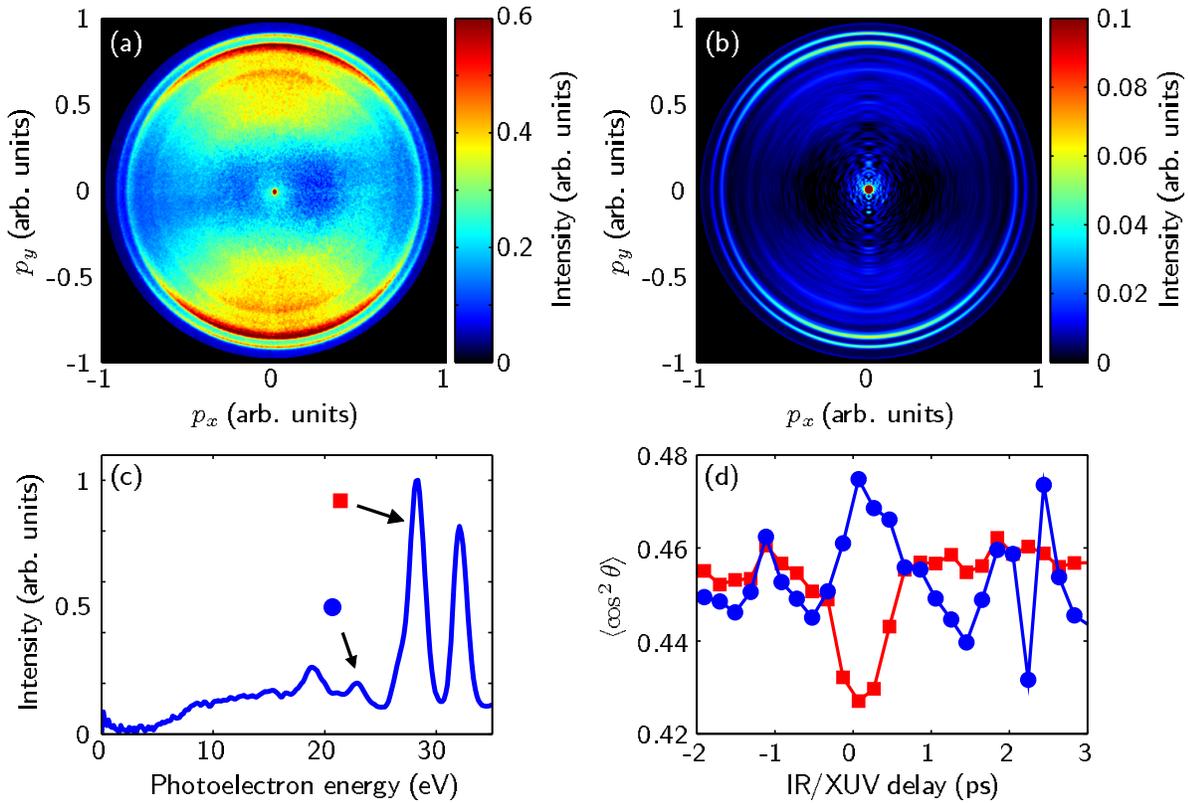}
\caption{\label{fig:mfpads}(Colour online) 2D Photoelectron momentum distributions resulting from ionization of CO$_{2}$ by the FEL (a), and a cut through the corresponding 3D momentum distribution (b). The $p_y$-axis is parallel to the XUV polarization axis. Panel (c) shows the corresponding photoelectron spectrum while panel (d) shows the delay-dependence of $\langle\cos^{2}\theta\rangle$ for the two photoelectron peaks indicated in (c), around the overlap between the FEL and the IR pulses.}
\end{figure}

\subsection{\label{ssec:MFPS}Towards molecular frame photoelectron imaging}
As discussed in the introduction, one of the main motivations for creating samples of aligned molecules at XUV (and future x-ray) FEL facilities like FLASH, is the prospect of performing high signal experiments directly in the molecular frame. This would enable for example electron diffraction experiments where the molecules are \textquotedblleft illuminated from within", and the available short pulse durations would in addition allow to resolve ultrafast structural molecular dynamics in time.

A first attempt to assess the effect of molecular alignment on the photoelectron angular distribution was performed in the current experiments, with the results summarized in \fref{fig:mfpads}. The recorded 2D photoelectron distribution from ionization of CO$_{2}$ by the FEL is represented in \fref{fig:mfpads}(a) whereas a cut through the 3D distribution resulting from inversion of the data is shown in \fref{fig:mfpads}(b). The $p_y$-axis is parallel to the XUV polarization axis. The corresponding kinetic energy spectrum is shown in \fref{fig:mfpads}(c). The images show multiple rings, corresponding to different ionization pathways whose assignment to the different states can be found in~\cite{JohnssonJMO2008}. In two of the rings weak time-dependent effects in the angular distribution are observed. Panel (d) shows the $\langle\cos^{2}\theta\rangle$-values obtained for the two rings around the time overlap between the FEL and the IR pulses. In particular for the strongest of the two peaks (red squares), the signal-to-noise ratio is high and a clear decrease of the $\langle\cos^{2}\theta\rangle$-value can be seen. The weakest peak (blue circles) shows an opposite behavior, although the observation is somewhat hampered by the high noise level. The first peak has a kinetic energy distribution that is centered around 28.2~eV [see \fref{fig:mfpads}(d)]. It can be assigned to a multiple contribution from the A$^{2}\Pi^{+}_{u}$ and B$^{2}\Sigma^{+}_{u}$ state present in this binding energy region. The asymmetric feature of this ring around E$_{k}$=26.5~eV was associated in our previous study with ionization to the C$^{2}\Sigma^{+}_{g}$ state \cite{JohnssonJMO2008}. The second peak is found around E$_{k}$=22.9~eV and has been designated as originating from a multiple electron transition~\cite{BrionCP1978,LochtIHMSIP1995}. Since several states contribute to each of the rings (that could not be resolved in this measurement), and since the effect that was observed is very weak, further analysis is very difficult. However, recently we have repeated the experiment using our newly built attosecond setup at AMOLF where a molecular beam of CO$_{2}$ was first exposed to a femtosecond near-IR aligning pulse and later probed by a femtosecond XUV pulse, produced from high-order harmonic generation in argon. The preliminary analysis of the photoelectron angular distributions reveals a very clear dependence on the induced molecular alignment when the pulses overlap, and also in the field-free case, around the revivals in the alignment. The effect is very similar to the one observed in this study, and a closer investigation is under way, aiming at fully understanding these results.

\section{\label{sec:outlook}Conclusion \& outlook}
In conclusion, we have shown that field-free molecular alignment can be induced and probed at FLASH, the free electron laser in Hamburg. By using higher IR field intensities, possibly in combination with pulse shaping techniques~\cite{HertzPRA2007,SuzukiPRL2008}, one can expect the degree of alignment to be further improved, giving opportunities to perform molecular frame experiments including electron spectroscopy and electron diffraction measurements.

\ack This work is part of the research program of the \textquotedblleft Stichting voor Fundamenteel Onderzoek der Materie (FOM)", which is financially supported by the \textquotedblleft Nederlandse organisatie voor Wetenschappelijk Onderzoek (NWO)". Financial support by the Marie Curie Research Training Networks 'XTRA' is gratefully acknowledged. Portions of this research were carried out at the light source facilities FLASH at HASYLAB/DESY. DESY is a member of the Helmholtz Association (HGF). We are greatly indebted to the scientific~\cite{AckermannNPhot2007} and technical team at FLASH, in particular the machine operators and run coordinators, being the foundation of the successful operation and delivery of the SASE-FEL beam. We also acknowledge travel support from the European Community under Contract RII3-CT-2004-506008 (IA-SFS). P.~J. acknowledges the support of the Swedish Research Council, and M.~F.~K. is grateful for support by the Emmy-Noether program of the DFG. Finally, Rob Kemper, Ad de Snaijer and Hinco Schoenmaker are gratefully acknowledged for their work in preparing and installing the experimental apparatus at FLASH.


\providecommand{\newblock}{}

\end{document}